\documentclass[a4paper]{article}

\usepackage{INTERSPEECH2019}
\usepackage[utf8]{inputenc} 
\usepackage[T1]{fontenc} 
\usepackage{tikz}
\usepackage{pgfplots}
\usepackage{array}

\newcolumntype{P}[1]{>{\centering\let\newline\\\arraybackslash\hspace{0pt}}m{#1}}

\title{Real to H-space Encoder for Speech Recognition}
\name{Titouan Parcollet$^{1,2}$, Mohamed Morchid$^1$, Georges Linarès$^1$, Renato De Mori$^3$}
\address{
  $^1$Avignon Université, LIA, France \\
  $^2$ORKIS, Aix-en-provence, France \\
  $^3$McGill University, Montréal, QC, Canada}
\email{titouan.parcollet@alumni.univ-avignon.fr, \{firstname.lastname\}@univ-avignon.fr}

\begin{document}

\maketitle
\begin{abstract}
 Deep neural networks (DNNs) and more precisely recurrent neural networks (RNNs) are at the core of modern automatic speech recognition systems, due to their efficiency to process input sequences. Recently, it has been shown that different input representations, based on multidimensional algebras, such as complex and quaternion numbers, are able to bring to neural networks a more natural, compressive and powerful representation of the input signal by outperforming common real-valued NNs. Indeed, quaternion-valued neural networks (QNNs) better learn both internal dependencies, such as the relation between the Mel-filter-bank value of a specific time frame and its time derivatives, and global dependencies, describing the relations that exist between time frames. Nonetheless, QNNs are limited to quaternion-valued input signals, and it is difficult to benefit from this powerful representation with real-valued input data. This paper proposes to tackle this weakness by introducing a real-to-quaternion encoder that allows QNNs to process any one dimensional input features, such as traditional Mel-filter-banks for automatic speech recognition.    
  
\end{abstract}

%
%
\section{Introduction}
\label{sec:intro}

Automatic speech recognition (ASR) systems have been widely impacted by machine learning, and more precisely by the resurgence of deep neural networks (DNNs). In particular, recurrent neural networks (RNNs) have been designed to learn parameters of sequence to sequence mapping, and various models were successfully applied to ASR with a remarkable increasing in the ASR system performance. In order to avoid parameter estimation problems, RNNs with long short-term memory \cite{hochreiter1997long,schuster1997bidirectional}, and gated recurrent unit (GRU) \cite{chorowski2015attention} have been proposed to mitigate vanishing and exploding gradients when learning long input sequences. Nevertheless, less attention has been paid to model input features with multiple views of speech spectral tokens.

A noticeable exception is the use of complex-valued numbers in neural networks (CVNNs) to jointly represent amplitude and phase of spectral samples \cite{chiheb2017complex}. More recently, quaternion-valued neural networks (QNNs) have been investigated to process the traditional Mel-frequency cepstral coefficients (MFCCs), or Mel-filter-banks plus time derivatives \cite{parcollet2019qlstm,parcollet2018qrnn,parcollet2018qcnn} as composed entities. Superior accuracy, with up to four times less model parameters, has been observed with quaternion-valued models compared to results obtained with real-valued equivalent models. In fact, common real-valued neural networks process energies and time derivatives independently, learning both global dependencies between multiple time frames, and local
values in a specific time frame, without considering the relations between a value and its derivatives. Instead, the quaternion algebra allows QNNs \cite{parcollet2019qlstm,parcollet2018qrnn,nitta1995quaternary,arena1997multilayer,isokawa2009quaternionic,chase2017quat} to process time frames as composed entities, with internal relations learned within the specific algebra, and global dependencies learned with the neural network architecture, while reducing by an important factor the number of neural parameters needed. As a side effect, with quaternion algebra, the number of neural parameters to be estimated is reduced by an important factor. Nonetheless, QNNs input features must be encoded as quaternion numbers, requiring a preliminary definition of input views that cannot be modified by the learning process. In many cases, it looks advantageous to have multiple input feature views, but there may be different choices of them and it is not clear how to make a selection. Examples could be views based on temporal or spectral relations. In this paper, a real-to-quaternion encoder (R2H) is proposed to let a quaternion-valued neural architecture learn hidden representations of input feature views. The R2H layer acts as an encoder to train QNNs with any real-valued input vector. Indeed, this encoder allows the model to learn in an end-to-end architecture, a latent and quaternion-valued representation of the input data. This representation is then used as an input to a quaternion-valued classifier, to exploit the capabilities of quaternion neural networks. For achieving this objective, the contributions of this paper are:
\begin{itemize}
    \item Investigate different real-to-quaternion (R2H) encoders to learn an internal representation of any real-valued input data (Section \ref{sec:project}). 
    \item Merge the R2H with the previously introduced quaternion long short-term memory neural network (QLSTM, Section \ref{sec:qlstm})\cite{parcollet2018qrnn}\footnote{Code is available at: https://github.com/Orkis-Research/Pytorch-Quaternion-Neural-Networks}.
    \item Evaluate this approach on the TIMIT, and Librispeech speech recognition tasks (Section \ref{sec:exps}).
\end{itemize}
Results improvements on both TIMIT and Librispeech speech recognition tasks are reported with the introduction of a R2H encoder in a QLSTM architecture, with input 
made of $40$ Mel-filter-bank coefficients, and with more than three times
fewer neural parameters than with real-valued LSTMs.

%
%
\section{Quaternion algebra}
\label{sec:qalgebra}

The quaternion algebra $\mathbb{H}$ defines operations between quaternion numbers. A quaternion Q is an extension of a complex number to the hyper-complex plane defined in a four dimensional space as:
\begin{align}
Q = r1 + x\textbf{i} + y\textbf{j} + z\textbf{k},
\end{align}
where $r$, $x$, $y$, and $z$ are real numbers, and $1$, \textbf{i}, \textbf{j}, and \textbf{k} are the quaternion unit basis. In a quaternion, $r$ is the real part, while $x\textbf{i}+y\textbf{j}+z\textbf{k}$ with $\textbf{i}^2=\textbf{j}^2=\textbf{k}^2=\textbf{i}\textbf{j}\textbf{k}=-1$ is the imaginary part, or the vector part. 
Such a definition can be used to describe spatial rotations. A quaternion $Q$ can also be summarized into the following matrix of real numbers, that turns out to be more suitable for computations:
\begin{align}
\label{eq:qmat}
Q_{mat} = 
\begin{bmatrix}
   r & -x & -y & -z \\
   x & r & -z & y \\
   y & z & r & -x \\
   z & -y & x & r 
\end{bmatrix}.
\end{align}
The conjugate $Q^*$ of $Q$ is defined as:
\begin{align}
\label{eq:conjugate}
Q^*=r1-x\textbf{i}-y\textbf{j}-z\textbf{k}.
\end{align}
Then, a normalized or unit quaternion $Q^\triangleleft$ is expressed as:
\begin{align}
\label{eq:normalize}
Q^\triangleleft=\frac{Q}{|Q|},
\end{align}
with $|Q|$ the norm of Q defined as: 
\begin{align}
\label{eq:norm}
|Q|=\sqrt{r^2+x^2+y^2+z^2}.
\end{align}
Finally, the Hamilton product $\otimes$ between two quaternions $Q_1$ and $Q_2$ is computed as follows: 
\begin{align}
Q_1 \otimes Q_2=&(r_1r_2-x_1x_2-y_1y_2-z_1z_2)+\nonumber \\
			&(r_1x_2+x_1r_2+y_1z_2-z_1y_2) \boldsymbol i+\nonumber \\
            &(r_1y_2-x_1z_2+y_1r_2+z_1x_2) \boldsymbol j+\nonumber \\
            &(r_1z_2+x_1y_2-y_1x_2+z_1r_2) \boldsymbol k.
\label{eq:hamilton}
\end{align}
The Hamilton product is used in QNNs to perform transformations of vectors representing quaternions, as well as scaling and interpolation between two rotations following a geodesic over a sphere in the $\mathbb{R}^3$ space as shown in \cite{minemoto2017feed}.

%
%
\section{Quaternion long short-term memory neural networks}
\label{sec:qlstm}

Long short-term memory neural networks (LSTM) are a well-known and investigated extension of recurrent neural networks \cite{hochreiter1997long,greff2017lstm}. LSTMs offer an elegant solution to the vanishing and exploding gradient problems, alongside with a stronger learning capability of long and short-term dependencies within sequence. Following these strengths, a quaternion long short-term memory neural network (QLSTM) has been proposed \cite{parcollet2019qlstm}. 

In a quaternion-valued layer, all parameters are quaternions, including inputs, outputs, weights and biases. The quaternion algebra is ensured by manipulating matrices of real numbers \cite{parcollet2018qcnn,chase2017quat} to reconstruct the \textit{Hamilton product} from quaternion algebra. Consequently, for each input vector of size $N$, output vector of size $M$, dimensions are split in four parts: the first one equals to $r$, the second to $x$\textbf{i}, the third one is $y$\textbf{j}, and the last one equals to $z$\textbf{k}. The inference process of a fully-connected layer is defined in the real-valued space by the dot product between an input vector and a real-valued $M \times N$ weight matrix. In a QLSTM, this operation is replaced with the \textit{Hamilton product} '$\otimes$' (Eq. \ref{eq:hamilton}) with quaternion-valued matrices (\textit{i.e.} each entry in the weight matrix is a quaternion). 

Both LSTM and QLSTM networks rely on a gate action \cite{danihelka2016associative}, that allows the cell-state to retain or discard information from the past, and the future in the case of a bidirectional (Q)LSTM. Gates are defined in the quaternion space following \cite{parcollet2019qlstm}. Indeed, the gate mechanism implies a component-wise product of the components of the quaternion-valued signal with the gate potential in a split manner \cite{xu2017learning}.  Let $f_t$,$i_t$, $o_t$, $c_t$, and $h_t$ be the forget, input, output gates, cell states and the hidden state of a QLSTM cell at time-step $t$. QLSTM equations are defined as:
\begin{align}
    f_t =& \sigma(W_{f} \otimes x_t + R_{f} \otimes h_{t-1} + b_f),\\
    i_t =& \sigma(W_{i} \otimes x_t + R_{i} \otimes h_{t-1} + b_i),\\
    c_t =& f_t\times c_{t-1} + i_t\times \alpha(W_{c} \otimes x_t + R_{c} \otimes h_{t-1} + b_c),\\
    o_t =& \sigma(W_{o} \otimes x_t + R_{o} \otimes h_{t-1} + b_o),\\
    h_t =& o_t \times \alpha(c_t),
\end{align}
with $\sigma$ and $\alpha$ the Sigmoid and Tanh \textit{quaternion split activations} \cite{xu2017learning,arena1997multilayer}. 

Bidirectional connections allow (Q)LSTM networks to consider the past and future information at a specific time step, enabling the model to capture a more global context \cite{schuster1997bidirectional}. Quaternion bidirectional connections are identical to real-valued ones. Indeed, past and future contexts are added together component-wise at each time step.

An adapted scheme initialization for quaternion neural networks parameters has been proposed in \cite{parcollet2018qrnn}. In practice, biases are set to zero while weights are sampled following a Chi-distribution with four degrees of freedom. Finally, QRNNs require a specific backpropagation algorithm detailed in \cite{parcollet2018qrnn}.

\begin{figure*}
    \centering
    \includegraphics[scale=0.032]{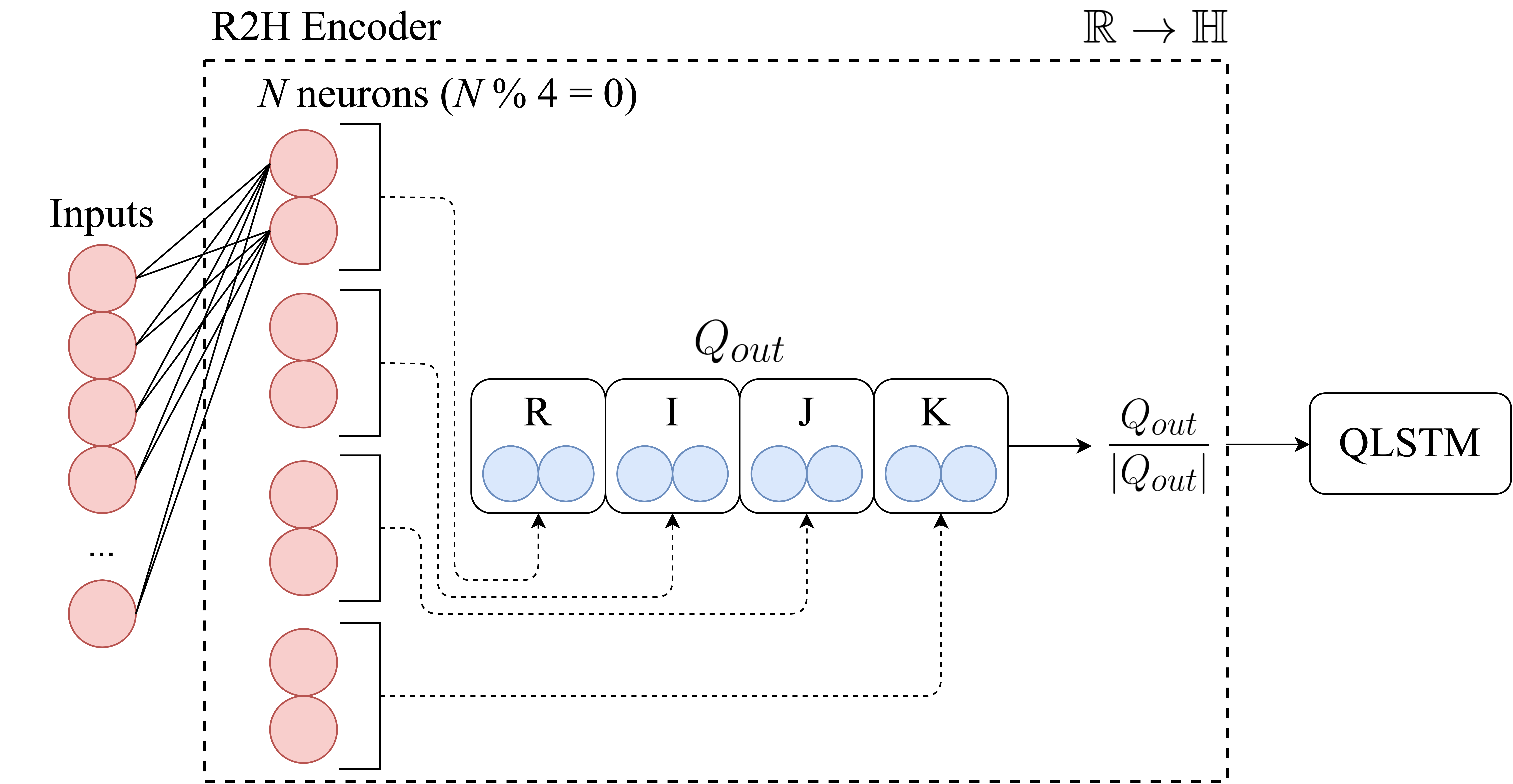}
    \caption{Illustration of the R2H encoder, used as an input layer to a QLSTM. Inputs are real, before being turned into quaternions, and finally unitary quaternions within the R2H encoder.}
    \label{fig:model}
\end{figure*}

%
%
\section{R2H encoder}
\label{sec:project}

As mentioned in the introduction, having input features represented by quaternions requires to have predefined a number of views for the same input token. This prevents the use of quaternion networks when prior knowledge suggests to use multiple views whose number and type cannot be exactly defined. For example, it is known that time relations between a speech spectrum and its neighbor spectra may improve the classification of the phoneme whose utterance produced the spectrum. Nevertheless, these relations may not be limited to time derivatives of all spectra samples in a spoken sentence. To overcome this limitation, a new method is proposed. It consists in introducing a real-valued encoder directly connected to the real-valued input signal. The real-to-quaternion (R2H) encoder is trained jointly to the rest of the model in an end-to-end manner, such as any other layer. After training, the encoder is expected to allow a mapping from the real space of the input features, to a latent internal representation meaningful for the following quaternion layers. The trained model is thus able to directly deal with real-valued input features, while internally processing quaternion numbers. The R2H encoder is a traditional dense layer followed by a quaternion activation function and normalization. The number of neurons contained in the layer must be a multiple of four for the quaternion representation. Let $W$, $X$ and $B$ be the weight matrix, the real-valued input and the bias vectors respectively. $Q^\triangleleft_{out}$ is the unit quaternion vector obtained at the output of the projection layer and is expressed as:

\begin{align}
Q^\triangleleft_{out}=\frac{Q_{out}}{|Q_{out}|},
\end{align}
with,
\begin{align}
Q_{out} = \alpha(W.X+B),
\end{align}
and $\alpha$ any quaternion split activation function. In practice, $Q_{out}$ and $Q^\triangleleft_{out}$ follow the quaternion internal representation defined in Section \ref{sec:qlstm}. Consequently, the input is split in four features from a latent sub-space, that are interpreted as quaternion components: the first one equals to $r$, the second to $x$\textbf{i}, the third one is $y$\textbf{j}, and the last one equals to $z$\textbf{k}, making it possible to apply the quaternion normalization, and the activation function. At the end of training $Q^\triangleleft_{out}$ capture an internal latent mapping of the real-valued input signal $X$ trough a vector of unit quaternions. Adding the R2H encoder as an input layer to QLSTMs or to any other QNNs allow the model to deal with real-valued inputs, while taking the strengths of QNNs (Figure \ref{fig:model}).

%
%
\section{Experiments}
\label{sec:exps}

Model architectures used for the experiments are presented in Section \ref{subsec:models}. Then, the R2H encoder is compared to the traditional and naive quaternion representation on the TIMIT and Librispeech speech recognition tasks (Section \ref{subsec:results})

\subsection{Model architectures}
\label{subsec:models}

QLSTMs have already been investigated for speech recognition in \cite{parcollet2019qlstm} and \cite{parcollet2018qrnn}. Consequently, and based on these previous researches, QLSTMs are composed with four bidirectional QLSTM layers with an internal real-valued size of $1,024$, equivalent to $256$ quaternion neurons. Indeed, $256 \times 4 = 1,024$ real numbers. The R2H encoder size varies from $256$ to $1,024$ to explore the best latent quaternion representation. Tanh, HardTanh and ReLU activation functions are investigated to compare the impact of bounded (Tanh, Hardtanh) and unbounded (ReLU) R2H encoders. In fact, the quaternion normalization allows a numerical reduction of the internal representation, but the ReLU counteracts the latter effect by integrating high real and positive values to the encoding. The final layer is real-valued and corresponds to the HMM states obtained with the Kaldi \cite{Povey_ASRU2011} toolkit. A dropout of $0.2$ is applied across all the layers except the output. The Adam optimizer \cite{kingma2014adam} is used to train the models with vanilla hyperparameters. The learning rate is halved every-time the loss on the validation set is below a certain threshold fixed to $0.001$ to avoid overfitting. Finally, models are implemented with the Pytorch-Kaldi toolkit \cite{pytorch-kaldi}. While the effectiveness of QLSTM over LSTM has been demonstrated, an LSTM network trained in the same conditions and based on \cite{parcollet2019qlstm} is considered as a baseline. All the models are trained during $30$ epochs, and the results on both the validation and test sets are saved at this point. 

\subsection{Phoneme recognition with the TIMIT corpus}
\label{subsec:results}

The training process is based on the standard $3,696$ sentences uttered by $462$ speakers, while testing is conducted on $192$ sentences uttered by $24$ speakers of the TIMIT \cite{garofolo1993darpa} dataset. A validation set composed of $400$ sentences uttered by $50$ speakers is used for hyper-parameter tuning. The raw audio is processed with a window of $25$ms and an overlap of $10$ms. Then, $40$-dimensional log Mel-filter-bank coefficients are extracted with the Kaldi toolkit. In previous work with QLSTMs \cite{parcollet2018qrnn,parcollet2019qlstm}, first, second and third time order derivatives were composed with spectral energies to build a multidimensional input representation with quaternions. In this paper, the time derivatives are no longer used. Instead, latent representations are directly learned from the R2H encoder, fed with the $40$ log Mel-filter bank coefficients. For the sake of comparison, an input quaternion is naively composed with input features from four consecutive Mel-filter-bank coefficients, before being fed to a standard QLSTM.

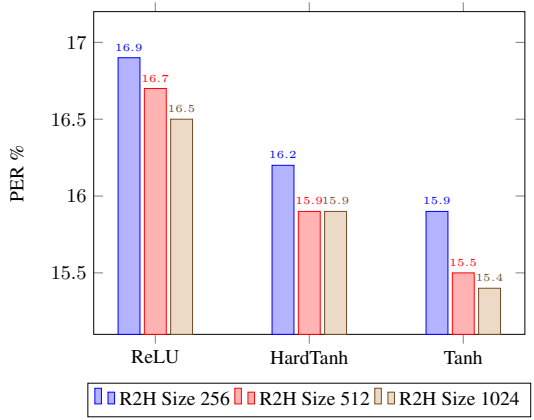
\begin{figure}[!h]
\centering
\scalebox{0.82}{
\begin{tikzpicture}
\begin{axis}[
    width=0.5\textwidth,
    height=.4\textwidth,
    ybar,
    enlargelimits=0.2,
    ylabel={PER \%},
    symbolic x coords={ReLU, HardTanh, Tanh},
    xtick=data,
    legend style={at={(0.5,-0.15)},
    anchor=north,legend columns=-1},
    nodes near coords,
    every node near coord/.append style={font=\tiny},
    nodes near coords align={vertical},
    ]
\addplot coordinates {(ReLU, 16.9) (HardTanh, 16.2) (Tanh, 15.9)};
\addplot coordinates {(ReLU,  16.7) (HardTanh, 15.9) (Tanh, 15.5) };
\addplot coordinates {(ReLU, 16.5) (HardTanh, 15.9) (Tanh, 15.4) };
\legend{R2H Size 256,  R2H Size 512,  R2H Size 1024}
\end{axis}
\end{tikzpicture}
}
\caption{Phoneme Error Rate (PER \%) obtained on the test set of the TIMIT corpus with different activation functions, and different R2H encoder size for a QLSTM. Results are from an average of three runs.}
\label{fig:results}
\end{figure}

Figure \ref{fig:results} reports the results obtained for the investigation of the R2H encoder size and the impact of the activation layer. Results are from an average of three runs and are not obtained w.r.t to the validation set. Indeed, performances on the test set are evaluated only once at the end of the training phase. It is first interesting to note that a layer of $1,024$ neurons always gives better results than a layer of size $256$ or $512$, without even considering activation functions. In the same manner, the Tanh activation outperforms both ReLU and Hardtanh activation function with all the layer size, with an average phoneme error rate (PER) on the TIMIT test set of $15.6$\% compared to $16.7$ and $16$\% for the ReLU and HardTanh activations. It is important to note that the ReLU activation gives the worst results. An explanation of such phenomenon is the definition interval of the ReLU function. When dealing with ReLU, outputs of the R2H layer are not bounded in the positive domain before being normalised. Therefore, the dense layer can output large values that are then squashed by the quaternion normalization, and it can be hard for the neural network to learn such mapping. Conversely, both Hardtanh and Tanh functions are bounded by $-1$ and $1$, making it easier to learn, since values of the R2H layer before and after normalization vary on the same range. The HardTanh function also hardly saturates at $-1$ and $1$ in the same manner as the ReLU activation for negative numbers, while the Tanh smoothly tends to these bounds. Consequently, the HardTanh gives slightly worst results than the Tanh. Finally, a best PER of $15.4$\% is obtained with a normalised R2H encoder of size $1,024$ based on the Tanh activation function, compared to $16.5$\% and $15.9$\% with ReLU and Hardtanh functions.

Table \ref{table:results1} presents a summary of the results observed on the TIMIT phoneme recognition task with a QLSTM and basic quaternion features, compared to the proposed QLSTM coupled with the best R2H encoder from Fig. \ref{fig:results}. For fair comparison, a real-valued LSTM is also tested. As highlighted in \cite{parcollet2018qrnn}, QLSTMs models require less neural parameters than LSTMs due to their internal quaternion algebra. Therefore, an LSTM with $1,024$ neurons per layer is composed of $46.0$ million parameters, while corresponding QLSTMs only need $15.5M$ parameters. It is first interesting to note that the R2H encoder helps the QLSTM to obtains the same PER as the real-valued LSTM, while dealing with a real-valued input signal. Indeed, both models performed at $15.4$\% on the test set, while the QLSTM still requires more than three times fewer neural parameters.

\begin{table}[h!]
\centering
\caption{Phoneme error rate (PER\%) of the models on the development and test sets of the TIMIT dataset. “Params" stands for the total number of trainable parameters. "R2H-Norm" and "R2H" correspond to R2H encoders with and without normalization. Results are from an average of $3$ runs.}
\scalebox{0.9}{
    \begin{tabular}{ P{2.8cm}P{1.3cm}P{1.3cm} P{1.2cm}}
        \hline\hline
        \textbf{Models}& \textbf{Dev.}& \textbf{Test}& \textbf{Params}\\
        \hline
        
            LSTM& 14.5  & 15.4 & 46.0M \\
            QLSTM& 14.9 & 15.9 & 15.5M \\
            \hline
           	R2H-QLSTM & 14.7 & 15.7 & 15.5M  \\
            R2H-Norm-QLSTM & 14.4 & 15.4 &  15.5M \\
        \hline
    \end{tabular}
    \label{table:results1}
}
\end{table}

Then, it is worth underlining that the basic QLSTM without R2H layer obtains the worst PER of all models with $15.9$\% on the test set, due to the inappropriate input representation. Then, the impact of the quaternion normalization process is investigated by comparing a R2H encoder without normalization, to a normalised one. As expected, the quaternion normalization helps the input to fit the quaternion representation, and thus gives better results with $15.4$\% of PER compared to $15.7$\% for the non-normalized R2H encoder. It is important to mention that such results are obtained without batch-normalization, speaker adaptation or rescoring methods. 

\subsection{Speech recognition with the Librispeech corpus}

The experiments are extended to the larger Librispeech dataset \cite{panayotov2015librispeech}. Librispeech is composed of three distinct training subsets of $100$, $360$ and $500$ hours of speech respectively, representing a total training set of $960$ hours of read English speech. In our experiments, the models are trained following the setups described in Section \ref{subsec:models}, and based on the \textit{train\_clean\_100} subset containing $100$ hours. Results are reported on the \textit{test\_clean} set. Input features are the same as for the TIMIT experiments, and the best activation function reported in Figure \ref{fig:results} is used (Tanh). No regularization techniques such as batch-normalization are used, and no rescoring methods are applied at testing time. 

\begin{table}[h!]
\centering
\caption{Word error rate (WER\%) of the models on test\_clean set of the Librispeech dataset with a training on the train\_clean\_100 subset. “Params" stands for the total number of trainable parameters. "R2H-Norm" and "R2H" correspond to R2H encoders with and without normalization. No rescoring technique is applied.}
\scalebox{0.9}{
    \begin{tabular}{ P{2.8cm}P{1.3cm} P{1.3cm}}
        \hline\hline
        \textbf{Models}&  \textbf{Test}& \textbf{Params}\\
        \hline
        
            LSTM&  8.1 & 49.0M \\
            QLSTM&  8.5 & 17.7M \\
            \hline
           	R2H-QLSTM& 8.3 & 17.7M  \\
            R2H-Norm-QLSTM& 8.0 & 17.7M \\
        \hline
    \end{tabular}
    \label{table:results2}
}
\end{table}

The total number of neural parameters is slightly different when compared with the TIMIT experiments due to the increased number of HMM states, and therefore neurons, of the output layer for the Librispeech task. Nonetheless, the number of parameters is still lowered by a factor of $3$ when using QLSTM networks, compared to the real-valued LSTM. Similarly to the TIMIT experiments, the QLSTM with a normalized R2H layer reaches slightly better performances in term of word error rate (WER), with $8.0$\% compared to $8.1$\% for the LSTM. Moreover, the R2H encoder allows the QLSTM WER to decrease from $8.5$\% to $8.0$\%, representing an absolute gain of $0.5$\%. The reported results on the larger Librispeech dataset demonstrate that the R2H encoder solution scales well with more realistic speech recognition tasks.   

%
%
\section{Conclusions}

\textbf{Summary.} This paper addresses one of the major weakness of quaternion-valued neural networks known as the inability to process non quaternion-valued input signal. A new real-to-quaternion (R2H) encoder is introduced, making it possible to learn in a end-to-end manner a latent quaternion representation from any real-valued input data. Such representation is then processed with QNNs such as a quaternion LSTM. The experiments conduced on the TIMIT phoneme recognition task demonstrate that this new approach outperforms a naive quaternion representation of the input signal, enabling the use of QNNs with any type of inputs.

\noindent\textbf{Future work.} 
Split activation functions and current quaternion gate mechanisms do not fully respect the quaternion algebra by considering each elements as uncorrelated components. A future work will consist on the investigation of purely quaternion recurrent neural networks, involving well-adapted activation functions, and proper quaternion gates.

\bibliographystyle{IEEEtran}




\end{document}